\documentstyle[aps]{revtex}

\begin{document}
\author{Alberto Garc\'\i a\thanks{
On sabbatical leave in 1998-1999 from Departamento de F\'{\i}sica
CINVESTAV--IPN. } \thanks{ E-mail address:
aagarcia@fis.cinvestav.mx}}
\address{Departamento de F\'{\i}sica, Facultad de Ciencia,\\
Universidad de Santiago de Chile, Avda. Ecuador 3493, Casilla 307, Santiago,
Chile\\
and\\
Departamento de F\'{\i}sica CINVESTAV--IPN, Apartado
Postal 14--740, C.P. 07000, M\'exico, D.F. M\'exico}
\title{On the Rotating Charged BTZ Metric}
\date{\today}
\maketitle

\begin{abstract}
{\bf Abstract:} {It is shown that the charged non-diagonal BTZ
(2+1)-spacetime is not a solution of the Einstein-Maxwell field
equations with cosmological constant.\\ }\\ \\
{Keywords: 2+1 dimensions, black hole.}\\ PACS numbers:
04.20.Jb\\
\end{abstract}

\smallskip\ Ba\~nados et al.~\cite{Teitelboim1} in their elegant paper
presented black hole solutions to the Einstein--Maxwell field
equations in (2+1)-anti de Sitter spacetime, from now on I refer to
them as BTZ solutions, opening with this work an area of intensive
research~\cite {Carlip,Mann,Frolov}. The BTZ solutions possess
features inherent to the (3+1)-black holes, and it is believed that
(2+1)-gravity will provide new insights to a better understanding
of the physically relevant (3+1)-gravity. Nevertheless, the
reported charged generalization of their non-diagonal metric (2),
pag. 1850, occurs to be wrong, it does not fulfill the
Einstein--Maxwell equations.

One constructs a (2+1)-Einstein theory coupled with Maxwell electrodynamics
starting from the action
\begin{eqnarray}  \label{action}
S=\frac{1}{2\pi}\int \sqrt{-g} \left( R-2\Lambda- F_{ab}F^{ab} \right)\,d^3x,
\end{eqnarray}
Varying this action with respect to gravitational field one obtains the 
Einstein equations
\begin{eqnarray}
G_{ab} + \Lambda g_{a b}=2( F_{ac} F_{b}^{\,\,c}-\frac{1}{4}%
g_{ab}F_{cd}F^{cd} ),
\end{eqnarray}
while the variation with respect to the electromagnetic potential $A_{a}$
entering in $F_{ab}= A_{b,a} - A_{a,b}$, yields the electromagnetic field
equations
\begin{eqnarray}  \label{elec}
\nabla_{a} \left( F^{ab} \right)=0 .
\end{eqnarray}

The non-diagonal BTZ (2+1)- metric has the form
\begin{eqnarray}
ds^2=-N^2dt^2+\frac{dr^2}{N^2}+r^2(d\phi +N^\phi dt)^2,  \label{metrica}
\end{eqnarray}
where $N^2=N^2(r)$ and $N^\phi =N^\phi (r)$ are unknown functions of the
variable $r$. The contravariant metric components one can read from $%
\partial _s^2=g^{ab}\partial _a\partial _b$:
\begin{eqnarray}
\partial _s^2=-\frac 1{N^2}\partial _t^2+N^2{\partial _r}^2+(\frac 1{r^2}-%
\frac{N^{\phi 2}}{N^2})\partial _{\phi {^2}}+2\frac{N^\phi }{N^2}\partial
_t\partial _\phi .  \label{tetrad1}
\end{eqnarray}

I restrict the electric field, as in \cite{Teitelboim1}, to be $%
F_{rt}=-E(r)=A_{t,r}$, thus
\begin{eqnarray}
F_{ab}=E(r)\left( \delta _a^t\delta _b^r-\delta _a^r\delta _b^t\right) .
\label{tensorr}
\end{eqnarray}
hence the non-vanishing contravariant $F^{ab}$ components are
\begin{eqnarray}
F^{rt}=-E,\,\,F^{r\phi }=-E\,N^\phi .  \label{tensr}
\end{eqnarray}
The invariant $F$ then is given by $2F=-E^2(r)\longrightarrow L(F)=E^2/8\pi $%
.

The electromagnetic field equations (\ref{elec}), in the considered case,
reduce to ${\frac d{dr}}\left( \sqrt{-g}F^{rb}\right) =0$, and yield
\begin{eqnarray}
{\frac d{dr}}(r\,\,E)=0\longrightarrow rE(r)=Q=constant,  \label{mag2}
\end{eqnarray}
where $Q$ is a constant interpretable as the charge, together with
\begin{eqnarray}
{\frac d{dr}}(r\,\,E\,\,N^\phi )=0.
\label{mag3}
\end{eqnarray}

Substituting $E(r)=Q/r$ from(\ref{mag2}) into the second electromagnetic
equation (\ref{mag3}), one arrives at
\begin{eqnarray}
{\frac d{dr}}N^\phi (r)=0\longrightarrow N^\phi (r)=N_0^\phi =constant.
\label{bz2}
\end{eqnarray}

This means that the metric, instead of being non-diagonal is simply
static (diagonal); the metric reduces to the diagonal form by
introducing a new Killingian coordinate $d\phi ^{\prime }=d\phi
+N_0^\phi dt$, thus without loss of generality $N_0^\phi $ can be
equated to zero.\\A straightforward demonstration of the
incorrectness of the Ba\~nados et al charged solution can be
established as follows. If one substitutes into the Maxwell
equations above the functions presented in Ba\~nados paper, pag.
1850 third paragraph after eq.(6), for the charged case, namely:\
\begin{eqnarray}
A_{0}=-Q {\ln}(\frac{r}{r_{0}}),N^{\phi}=-\frac{J}{2r^{2}},
\end{eqnarray}
\begin{eqnarray}
{\vec E}=-(\vec{\nabla} A_{0})\rightarrow E=- {\frac{d}{dr}}
A_{0}=\frac{Q}{r},
\end{eqnarray}
where $J$ is the angular momentum constant, one concludes that the
equation for $E$,(\ref{mag2}), is readily fulfilled, while
(\ref{mag3})yields the condition
\begin{eqnarray}
\frac{QJ}{r^{3}}=0,
\end{eqnarray}
therefore in the charged case, $Q\neq0$, the angular momentum $J$
has to vanish, $J=0 $ !!, and hence $N^{\phi}=0$. One is then
arriving at a {\bf globally} diagonal metric ---a static charged
metric for the whole spacetime.

Although these approaches are sufficient to demonstrate the
incorrectness of the charged BTZ result, I shall prove this fact
also by using the Einstein- Maxwell equations:

\begin{eqnarray}
\frac 1{2r}{\frac d{dr}}\left( r{\frac d{dr}}N^2\right) -\frac 12(r{\frac d{%
dr}}N^\phi )^2=-2\Lambda ,  \label{riccit}
\end{eqnarray}
\begin{eqnarray}
\frac 1r{\frac d{dr}}N^2+\frac 12(r{\frac d{dr}}N^\phi )^2=-2E^2-2\Lambda ,
\label{ric2}
\end{eqnarray}
\begin{eqnarray}
3\,{\frac d{dr}}N^\phi +r{\frac{d^2}{d{r}^2}}N^\phi =0.
\label{rit}
\end{eqnarray}
The last equation above can be written as $(r^3N_{,r}^\phi
)_{,r}=0$ , thus it integrates as
\begin{eqnarray}
N_{,r}^\phi =\frac J{r^3}\longrightarrow N^\phi (r)=-\frac J{2r^2}+N_0^\phi ,
\label{rit2}
\end{eqnarray}
the shifting freedom of the Killingian ${\phi }$ coordinate can be
used to set the integration constant $N_0^\phi $ equal to zero.
Replacing the derivative of $N^\phi (r)$ into the first order
equation (\ref{ric2}) for $N^2(r)$, one has
\begin{eqnarray}
N_{,r}^2=-\frac{J^2}{2r^3}-2rE^2-2\Lambda r,  \label{rit5}
\end{eqnarray}
which integrates as
\begin{eqnarray}
N^2(r)=-M+\frac{J^2}{4r^2}-\Lambda r^2-2\int rE{^2}\,dr.  \label{rit6}
\end{eqnarray}
Let us now use the information contained in the Maxwell equations:\
\begin{eqnarray}
{\frac d{dr}}(r\,\,E)=0\longrightarrow rE(r)=Q=constant,
\end{eqnarray}
which substituted into $N^2(r)$ yields
\begin{eqnarray}
N^2(r)=-M+\frac{J^2}{4r^2}-\Lambda r^2-2Q^2{\ln r}.  \label{rit7}
\end{eqnarray}
At this stage, with $\Lambda =-1/l^2$, one recognize the structural
functions presented for the charged case in ~\cite{Teitelboim1}:
\begin{equation}
N^2=N_{Q=0}^2+\frac 12QA_0,\ N^\phi =-\frac J{2r^2},
\end{equation}
\begin{equation}
A_0=-Q{\ln }(\frac r{r_0}),\ E=-{\frac d{dr}}A_0.
\end{equation}
Finally, substituting $N^\phi (r)$ and $E(r)$ into the Maxwell equation for $%
N^\phi (r)$,
\begin{eqnarray}
{\frac d{dr}}(r\,\,E\,\,N^\phi )=0,
\end{eqnarray}
one arrives at the condition
\begin{eqnarray}
\frac{QJ}{r^3}=0,
\end{eqnarray}
Hence one has two possible cases:
\begin{enumerate}
\item[a)]  The electrovacuum plus lambda case with the charge different from
zero, $Q\neq 0$. In such a case $J=0$ and thus $N^\phi (r)$ has to be a
constant (reducible to zero), under this circumstances one obtains a correct
static charged (2+1)-solution with structural functions $N^2(r)=-M-\Lambda
r^2-2Q^2{\ln r}$, and $N^\phi (r)=0$. Thus one arrives at the main result of
this comment: {\em within the BTZ metric anzats (\ref{metrica}) there is no
charged non-diagonal solutions to the Einstein-Maxwell plus lambda equations
in (2+1)-dimensions}.

\item[b)]  The vacuum plus lambda case, with $Q=0$ and therefore $E=0$. In
this case the structural functions $N^2(r)=-M+J^2/4r^2-\Lambda r^2$ and $%
N^\phi (r)=-J/2r^2$ give rise to the non-diagonal metric reported in the BTZ
paper.
\end{enumerate}

The equation (\ref{riccit}) gives no further conditions.

To treat correctly the rotating charged plus lambda case, one has to enlarge
the class of metrics, namely to deal with a line element of the form
\begin{eqnarray}
\ ds^2=-fdt^2+\frac{dr^2}g+R^2(d\phi +Wdt)^2,
\end{eqnarray}
where $f=f(r)$, $g=g(r)$, $R=R(r)$ and $W=W(r)$ are unknown functions of the
variable $r$. One can use the freedom in the choice of the $r$ variable to
fix one of the functions above. The electromagnetic tensor can be assumed to
be of the form
\begin{eqnarray}
\ F_{ab}=E(r)({{\delta }^t}_a{{\delta }^r}_b-{{\delta }^r}_a{{\delta }^t}%
_b)+B(r)({{\delta }^\phi }_a{{\delta }^r}_b-{{\delta }^r}_a{{\delta }^\phi }%
_b).
\end{eqnarray}
When this work was in the refereeing process at PRL, the author learnt
from the referee that a footnote on this respect appeared in~\cite{Kamata}.

This work was supported in part by FONDECYT (Chile)-1980891,
CONACYT (M\'exico)-3692P-E9607.

\end{document}